\documentclass[fleqn,usenatbib]{mnras}

\usepackage{newtxtext,newtxmath}

\usepackage[T1]{fontenc}

\DeclareRobustCommand{\VAN}[3]{#2}
\let\VANthebibliography\thebibliography
\def\thebibliography{\DeclareRobustCommand{\VAN}[3]{##3}\VANthebibliography}

\usepackage{graphicx}	
\usepackage{amsmath}	
\usepackage[markup=nocolor,addedmarkup=bf]{changes} 

\graphicspath{{./}{figures/}}

\title[Post-Newtonian parameter and Hubble constant from strongly lensed FRBs]{Prospects of strongly lensed fast radio bursts: Simultaneous measurement of post-Newtonian parameter and Hubble constant}

\author[Gao et al.]{Ran Gao, Zhengxiang Li\thanks{E-mail: zxli918@bnu.edu.cn}, He Gao
\\
Department of Astronomy, Beijing Normal University, Beijing 100875, China; \\
}
\date{Accepted XXX. Received YYY; in original form ZZZ}

\pubyear{2021}

\begin{document}
\label{firstpage}
\pagerange{\pageref{firstpage}--\pageref{lastpage}}
\maketitle

\begin{abstract}
Strong gravitational lensing effect is a powerful tool to probe cosmological models and gravity theories. Recently, the time-delay cosmography from strong lensing and the stellar kinematics of the deflector, which encode the Hubble constant and the post-Newtonian parameter via two distance ratios reflecting the lensing mass and dynamical mass respectively, have been proposed to investigate these two parameters simultaneously. Among strong lensing systems with different sources, strongly lensed fast radio bursts (FRBs) have been proposed as precision probes of the universe since the time delay $\sim$ 10 days between images could be measured extremely precisely because of their short duration of a few milliseconds. In this work, we investigate the ability of strongly lensed FRBs on simultaneously estimating these two parameters via simulations. Take the expected FRB detection rate of upcoming facilities and lensing probability into consideration, it is likely to accumulate 10 lensed FRBs in several years and we find that $H_0$ could be determined to a $\sim1.5\%$ precision and $\gamma_{\rm PPN}$ could be constrained to a $\sim8.7\%$ precision simultaneously from them. These simultaneous estimations will be helpful for properly reflecting the possible correlation between these two fundamental parameters.

\end{abstract}

\begin{keywords}
gravitational lensing --- fast radio bursts --- cosmological parameters
\end{keywords}
 
\section{Introduction}\label{sec:intro}

General relativity (GR) is one of the most important and famous theories in the past century because it has been an extremely successful description of gravity. GR has passed almost all experimental tests, the first famous Eddington's light deflection measurement during the 1919 solar eclipse~\citep{1920RSPTA.220..291D}; gravitational redshift observation~\citep{1960PhRvL...4..337P}; Shapiro delay measurements~\citep{1964PhRvL..13..789S, 2003Natur.425..374B}; extensive studies of relativistic effects in binary radio pulsar systems, including verification of energy loss via gravitational waves (GWs) as observed in the Hulse–Taylor pulsar~\citep{1979Natur.277..437T}; the successful operation of the Global Positioning System~\citep{2002PhysToday.55.5.41}; precise measurement of the Earth-Moon separation as a function of time through lunar laser ranging techniques~\citep{2004PhRvL..93z1101W}; especially the successful direct detection of GW from the merge of binary of compact objects~\citep{2016PhRvL.116f1102A}. In addition to these validated tests, GR also had has further achievements in its applications to physical cosmology. That is, on the basis of GR and the generalized Copernican principle, the evolution of the universe as a function of cosmic time can be predicted from the knowledge of densities of its constituents. The more fundamental of the role playing in astrophysics and cosmology, GR should be tested more rigidly by various ways. The parameterized post-Newtonian (PPN) framework~\citep{1971ApJ...163..595T}, which is traditionally characterized by a measure of the amount of spatial curvature per unit mass and usually denoted by the scale-independent parameter $\gamma_{\rm PPN}$ with $\gamma_{\rm PPN}=1$ representing GR, provides a systematic and quantitative way to formulate and interpret tests of gravity. For instance, the magnitude deviating from GR has been widely tested on kiloparsec (kpc) scales with galaxy-galaxy strong gravitational lensing observations~\citep{2006PhRvD..74f1501B,2009arXiv0907.4829S,2010ApJ...708..750S,2017ApJ...835...92C,2018Sci...360.1342C,2019ApJ...880...50Y,2020PhRvD.101j3505D}. In these tests, $\gamma_{\rm PPN}$ was constrained to $0.997^{+0.037}_{-0.047}$ from a mass-selected sample of galaxy-scale strong gravitational lenses~\citep{2017ApJ...835...92C} and $0.97\pm0.09$ from an individual nearby lensing system, ESO 325-G004~\citep{2018Sci...360.1342C}.

The present rate of cosmic expansion, the Hubble constant $H_0$ relating redshift to distance and time, directly sets the size and age scale of the universe. As the most accessible parameter in the cosmological model, the value of $H_0$ can be directly determined with local measurements of distances and redshifts. Moreover, it can also be estimated from observations of the cosmic microwave background (CMB) at redshift $z\ge1100$ in a cosmological model. The first release of CMB data from the European Space Agency $Planck$ mission yielded $H_0=67.2\pm1.2~{\rm km/s/Mpc}$ in the context of $\Lambda$CDM~\footnote{The conventional units of $H_0$ will frequently be omitted in the rest of this paper for readability. $\Lambda$CDM refers to the standard cosmological model with the cosmological constant ($\Lambda$) and cold dark matter (CDM).}~\citep{2014A&A...571A..16P}, which showed a difference of 3$\sigma$ from the results of the second iteration of the Supernovae and $H_0$ for the Equation of State of dark energy (SH0ES) program, $73.8\pm2.4$~\citep{2011ApJ...730..119R} (hereafter R11) and firstly indicated the Hubble constant tension. In the following several years after the tension revealed, a great deal of efforts have been made for reanalysing the R11 data and essentially the same results have been obtained, with $H_0$ ranging from 72.5 to 76.0~\citep{2013MNRAS.434.2866F,2014MNRAS.440.1138E,2017MNRAS.471.2254Z}. The third iteration of SH0ES doubled the calibrator sample and refined distance estimates to anchors to reach a 2.4\% determination of $H_0 = 73.2\pm1.7$~\citep{2016ApJ...826...56R} (hereafter R16), $3.4\sigma$ greater than the refined value from $Planck+\Lambda$CDM of $66.9\pm0.6$~\citep{2016A&A...596A.107P}. Again, a great number of reanalyses of the R16 data with many variations were carried out and resulted in $H_0$ values ranging from 73 to 74 and uncertainties from 2\% to 2.5\%~\citep{2017JCAP...03..056C,2018MNRAS.476.3861F,2018MNRAS.477.4534F,2018A&A...609A..72D}. Three years later, observations of 70 long-period Cepheids in the LMC and improved distance estimates to the Large Magellanic Cloud~\citep{2019Natur.567..200P} and NGC4258~\citep{2019ApJ...886L..27R} resulted in $H_0 = 74.0\pm1.4$ which raised the discrepancy with $Planck+\Lambda$CDM to 4.4$\sigma$~\citep{2019ApJ...876...85R}. More recently, observations from the Hubble Space Telescope (HST) of Cepheid variables in the host galaxies of 42 Type Ia supernovae (SNe Ia) yields $H_0 = 73.3\pm1.0$~\citep{2021arXiv211204510R}, which is in 5$\sigma$ tension with the estimate of $H_0$ from Planck CMB observations under $\Lambda$CDM. In the past decades, measures of $H_0$ with local distance redshift observations roughly range from 70 to 75. However, those based on the pre-combination version of $\Lambda$CDM consistently range from 67 to 68. This emergent and substantial Hubble constant tension has stimulated numerous possible theoretical and observational explanations~\citep{2019NatAs...3..891V,2021CQGra..38o3001D,2021A&ARv..29....9S}. The source of this long-standing tension is still an open issue. At this crossroads, independent probes being able to determine $H_0$ with precision $<2\%$ would be helpful for coming to a robust conclusion~\citep{2019NatAs...3..891V}. For example, time-delay measurements of strong lensing systems are one of the most competitive probes~\citep{2010ARA&A..48...87T,2017ApJ...839...70L}. In this route, the $H_0$ Lenses in COSMOGRAIL's Wellspring (H0LiCOW) collaboration has obtained the most precise constraint on $H_0$, $73.3^{+1.7}_{-1.8}$, from a joint analysis of time-delay measurements of six strongly lensed quasars in the flat $\Lambda$CDM framework~\citep{2017MNRAS.465.4914B,2017MNRAS.468.2590S,2017MNRAS.465.4895W,2019MNRAS.484.4726B}.

There is a consensus that strong gravitational lensing systems are a powerful and complementary probe for constraining $H_0$ and testing gravity. However, these two aspects have been separately studied with the tool of strong lensing for a long time. It should be noted that these two issues are correlated when we use lensing systems at cosmological distances to test gravity because both background cosmology and effects of deviation from GR enter into the lens equation. Recently,~\citet{2020MNRAS.497L..56Y} first proposed to simultaneously constrain the Hubble constant and post-Newtonian parameter from strong lensing time-delay measurements. In that work, they obtained simultaneous constraints on $H_0$ and $\gamma_{\rm PPN}$ from the reanalysis of the four public lenses distance priors from the H0LiCOW collaboration, $H_0=73.65^{+1.95}_{-2.26}$ and $\gamma_{\rm PPN}=0.87^{+0.19}_{-0.17}$ favoring the local universe value and suggesting no deviation from GR, respectively. However, the constraints are limited by the $\ge1\%$ uncertainty in measuring quasar time-delay and $\ge2\%$ uncertainty in modeling the lens potential. One way to overcome these obstacles is to use a different source.

Fast radio bursts (FRBs) are very bright radio transient signals with durations of milliseconds. They have been detected from several hundred MHz to several GHz. Since the first FRB discovery in 2007, more than 600 bursts have been discovered~\citep{2007Sci...318..777L,2013Sci...341...53T,frbcollaboration2021chimefrb}.
For most FRBs, their dispersion measures (DMs) are significantly higher than the ones expected from the integrated electron column density within the Milky Way. Thess excesses are an evidence for extra-galactic origins of these mysteries. One of them, FRB 121102, shows a repeating feature~\citep{2016Natur.531..202S}. It was then successfully localized to a nearby dwarf galaxy at $z = 0.1927$~\citep{2017Natur.541...58C}. At this moment, about twenty FRBs, including both repeaters and apparently one-off bursts, have been localized to their host galaxies \footnote{http://frbhosts.org}. These localizations finally establish the cosmological origins of FRBs. Although the progenitor and radiation mechanism of the FRBs is uncertain, it has been widely proposed to use as a tool for constraining cosmological parameters~\citep{2014ApJ...788..189G,2014PhRvD..89j7303Z}.
Recently, \citet{2018NatCo...9.3833L} proposed that strongly gravitationally lensed FRBs as precision probes. Compared with traditional strongly lensed quasars, they have several compelling advantages, including short durations, possible precise localization, and no dazzling active galactic nuclei contamination. Later, this kind of system has been widely proposed for conducting cosmological and astrophysical studies~\citep{2021A&A...645A..44W,2021PhRvD.103f3017P,2021ApJ...916...70Z,2021PhRvD.104j3515A}. Here, we investigate the power of strongly gravitationally lensed FRB as a tool to simultaneously constrain \textit{$H_0$} and the post-Newtonian parameter.

This paper is organized as follows. We briefly review theories of cosmology, gravity, and gravitational lensing in Sec.~\ref{sec:method}. In Sec.~\ref{sec:Simulations and Results} we introduce materials for simulation and present results of simultaneous constraints on \textit{$H_0$} and the post-Newtonian parameter. Finally, we conclude and discuss in Sec.~\ref{sec:conclusions}.

\section{theoretical background}\label{sec:method}
In a homogeneous and isotropic universe, spacetime could be described by the FRWL metric,
\begin{equation}
\mathrm{d} s^{2}=-c^{2} \mathrm{~d} t^{2}+a^{2}(t)\left(\frac{\mathrm{d} r^{2}}{1-K r^{2}}+r^{2} \mathrm{~d} \Omega^{2}\right),
\end{equation}
where $c$ is speed of light, $K$ is the parameter describing the spatial curvature of the universe. In this work, we set $K=0$ and assume a spatially flat universe which is in good agreement with what obtained from the latest $Planck$ CMB observations.
Under this assumption, the proper distance from the object at redshift $z_1$ to the object at redshift $z_2$ is,
\begin{equation}
D\left(z_{1}, z_{2}\right)=\frac{c}{H_{0}} \int_{z_{1}}^{z_{2}} \frac{(1+z)^{-1} d z}{\sqrt{\Omega_{m}(1+z)^{3}+\Omega_{\Lambda}}}.
\end{equation}
Another useful distance is the angular diameter distance from the object at redshift $z_1$ to the object at redshift $z_2$,
\begin{equation}
D^{A}\left(z_{1}, z_{2}\right)=\frac{c / H_{0}}{1+z_{2}}\int_{z_{1}}^{z_{2}} \frac{d z}{\sqrt{\Omega_{m}(1+z)^{3}+\Omega_{\Lambda}}}.
\end{equation}

In the limit of the weak gravitational field, the space-time gauge is characterized by the Newtonian potential $\Psi$ and the spatial curvature potential $\Phi$.
\begin{equation}
d s^{2}=-\left(1+\frac{2 \Psi}{c^{2}}\right) c^{2} d t^{2}+a^{2}\left(1-\frac{2 \Phi}{c^{2}}\right) d \vec{x}^{2}
\end{equation}
The ratio of $\Phi$ and $\Psi$ is defined as $\gamma_{\mathrm{PPN}}$, which represents the spatial curvature generated per unit mass. In general relativity, $\gamma_{\mathrm{PPN}} = 1$ or $\Phi = \Psi$. 
Strong gravitational lensing systems are a promising probe to constrain the $\gamma_{\mathrm{PPN}}$ by confronting two mass measurements, i.e. the lensing mass obtained from the angle of deflection of light and the dynamical mass derived from the stellar kinematics of the lens galaxy. The Newtonian potential ($\Psi$) only affects non-relativistic ingredients, such as baryons and dark matters, while the angle of deflection of light is directly related to the Weyl potential ($\Psi_{+}=\frac{\Psi+\Phi}{2}=\left(\frac{1+\gamma_{\mathrm{PPN}}}{2}\right) \Psi$). Thus, comparison between the kinetic mass and the lensing mass is actually the same as the one between the Newtonian and Weyl potentials.
In the post-Newtonian framework, the deflection angle relates to the lensing mass in terms of,
\begin{equation}\label{eq5}
\alpha_{\mathrm{PPN}}(\theta)=\left(\frac{1+\gamma_{\mathrm{PPN}}}{2}\right) \frac{1}{\pi} \int_{\mathbb{R}^{2}} d^{2} \theta^{\prime} \frac{\Sigma\left(D_l \theta^{\prime}\right)}{\Sigma_{\mathrm{cr}}} \frac{\theta-\theta^{\prime}}{\left|\theta-\theta^{\prime}\right|^{2}},
\end{equation}
where $\theta$ is the position of the lens image, $\Sigma$ is the surface mass density, $D_l$ is the angular diameter distance from the observer to the lens, and $\Sigma_{\mathrm{cr}}=\frac{c^{2}}{4 \pi G} \frac{D_{s}}{D_{l} D_{ls}}$ is the critical surface mass density, which depends on the angular diameter distance between the source and the lens.
From Equation~\ref{eq5}, we can find that the effect of deviation from GR ($\gamma_{\mathrm{PPN}}$) and the background cosmology ($H_0$) are highly degenerate. That is, the change of deflection angle may arise from two sources, either a variation in $\gamma_{\mathrm{PPN}}$ or change in distances which directly correlates to the Hubble constant via $H_0^{-1}$.

Fortunately, measurements of time delay between different images in a strong lensing system are very helpful to break the above-mentioned degeneracy. For a strong gravitational lensing system with multiple images, the time delay between image A and B can be expressed as,
\begin{equation}
\Delta t_{\mathrm{A B}}=\left(1+z_{l}\right) \frac{\Delta \phi_{\mathrm{A B}}}{c} D_{\Delta t} \label{Dt},
\end{equation}
where $\Delta \phi_{\mathrm{AB}}$ is the Fermat potential difference between the two images, which can be further written as
\begin{equation}
\Delta \phi_{\mathrm{A B}}=\left[\phi\left(\theta_{\mathrm{A}}, \beta\right)-\phi\left(\theta_{\mathrm{B}}, \beta\right)\right],
\end{equation}
and
\begin{equation}
\phi(\theta, \beta)=\left[\frac{(\theta-\beta)^{2}}{2}-\psi(\theta)\right],
\end{equation}
where $\beta$ is the position of the source, $\psi(\theta)$ is the scaled gravitational potential at the image position. Moreover, $D_{\Delta t}$ is the so-called time delay distance and a combination of three distances,
\begin{equation}\label{dt}
D_{\Delta t}=\left(1+z_l\right) \frac{D_l D_s}{D_{ls}}, 
\end{equation}
where $D_l$, $D_s$ and $D_{ls}$ are the angular diameter distances from the observer to the lens, from the observer to the source, and from the lens to the source, respectively. This distance is inversely proportional to $H_0$ and is the first distance we need. It should be noted that he above relation, i.e. Equation~\ref{dt}, holds for all cosmological backgrounds and metric gravity theories.

Using the method presented by H0LiCOW~\citep{2019MNRAS.484.4726B}, we can obtain the following equations from the kinematic information of the lensed galaxy,
\begin{equation}\label{eq11}
\left(\sigma_{v}\right)^{2}=\frac{D_s}{D_{ls}} c^{2} J\left(\xi_{\text {lens}}, \xi_{\text {light}}, \beta_{\text {ani}}\right), 
\end{equation}
where $\beta_{\mathrm{ani}}(r) \equiv 1-\frac{\sigma_{t}^{2}}{\sigma_{r}^{2}}$ is the stellar distribution anisotropy. $\sigma_{r}$ is the radial dispersion and $\sigma_{t}$ is the tangential dispersion. The function $J$ captures all the model components computed from angles measured on the sky and the stellar orbital anisotropy distribution. This function also encompass all the ingredients for computing the velocity dispersion. It suggests that since the stellar dynamics is determined only by the Newtonian potential, this distance ratio is not affected by the PPN parameter~\citep{2020MNRAS.497L..56Y}. Therefore, we express the actual inferred lensing model parameters in the Fermat potential as $\xi_{ \mathrm{lens}}^{\prime}$. After including the effect of deviation from the GR, Equation~\ref{eq11} should be updated as,
\begin{equation}
\frac{2}{1+\gamma_{\mathrm{PPN}}} \frac{D_s}{D_{ls}}=\frac{\sigma_{v}^{2}}{c^{2} J\left(\xi_{ \mathrm{lens}}^{\prime}, \xi_{\mathrm{light}}, \beta_{\mathrm{ani}}\right)} \label{dd}
\end{equation}
Furthermore, by defining $D_l^{\prime}=\frac{1+\gamma_{\mathrm{PPN}}}{2} D_l$
and combining Equations (\ref{Dt}), (\ref{dt}), and (\ref{dd}), we can obtain
\begin{equation}
D_l^{\prime}=\frac{1}{1+z_l} \frac{c \Delta t_{\mathrm{AB}}}{\Delta \phi_{\mathrm{AB}}} \frac{c^{2} J\left(\xi_{\text {lens}}^{\prime}, \xi_{\text {light}}, \beta_{\text {ani}}\right)}{\sigma_{v}^{2}}. \label{Dd}
\end{equation}
This is the second distance, i.e. the kinetic distance, which is necessary for constraining the PPN parameter.

In conclusion, two distances, i.e. Equations~(\ref{dt}) and (\ref{Dd}), can be simultaneously derived from measurements including time delays, spectroscopies of the deflector, and lensing images. As proposed in~\citet{2020MNRAS.497L..56Y}, it is possible to obtain a joint bounds on $H_0$ and $\gamma_{\mathrm{PPN}}$ from posteriors of these two distances.

\section{Simulations and Results} \label{sec:Simulations and Results}
The main purpose of this work is to quantify the power of  upcoming strongly lensed FRBs on constraining $H_0$ and $\gamma_{\mathrm{PPN}}$ simultaneously. In order to achieve this goal, several determining factors, such as the redshift distribution of future FRBs, the lensing probability of a source at redshift $z_s$, and the uncertainty of each measurement, should be addressed. In this section, we first briefly clarify these factors one by one. Next, with these factors prepared, we investigate the power of upcoming strongly lensed FRBs on estimating $H_0$ and testing GR via Monte Carlo simulations.

\subsection{FRB redshift distribution} \label{subsec:redshift distribution}
\begin{figure}
    \includegraphics[width=\columnwidth]{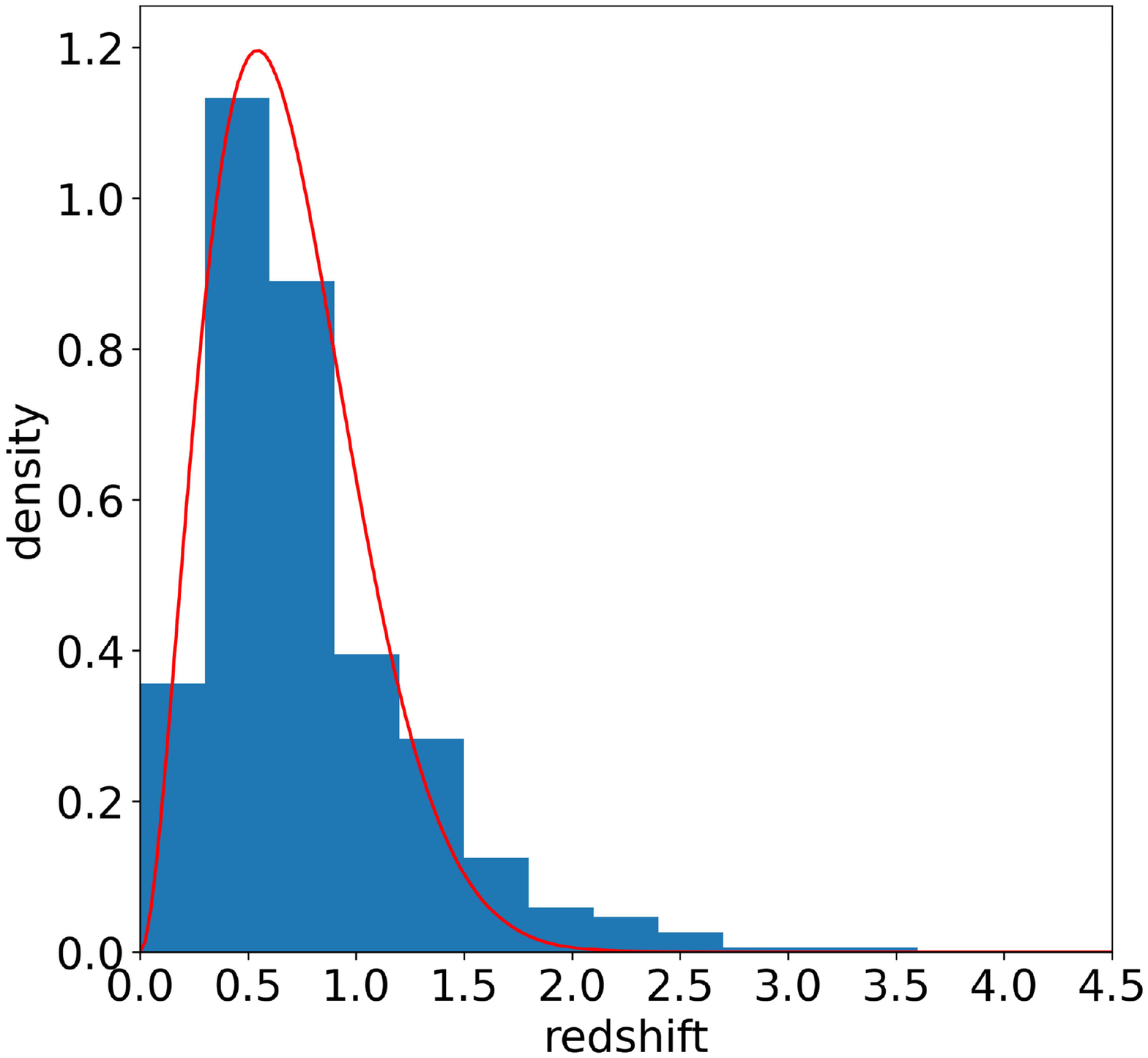}
    \caption{The number density with respect to redshift for $\sim600$ FRBs from the CHIME/FRB collaboration and other popular projects. The red curve is the result obtained by fitting the histogram with the distribution of Equation \ref{Nconst}. }
    \label{fig:1}
\end{figure}

Successful operations of several wide-field radio telescopes, such as the Canadian Hydrogen Intensity Mapping Experiment (CHIME), the Australian Square Kilometre Array Pathfinder (ASKAP), and the Deep Synoptic Array (DSA), have lead to a rapid increase of the number of detected FRBs. For instance, from 2018 July 25 to 2019 July 1, the CHIME/FRB Collaboration detected 535 FRBs including 61 bursts from 18 previously reported repeaters~\citep{2021arXiv210604352T}. All these FRBs together with bursts detected by other facilities have been collected and compiled by the Transient Name Sever (TNS)~\footnote{https://www.wis-tns.org}. So far, there are $>600$ independent events publicly available. Among these bursts, only a small fraction of them ($\sim 20$) have been localized to their host galaxy and thus their redshifts have been obtained. For the rest a large fraction of these bursts, we can roughly estimate their redshifts by using the relation $z \sim \mathrm{DM}_{\mathrm{E}} / 855~\mathrm{pc}~ \mathrm{cm}^{-3}$~\citep{2018ApJ...867L..21Z,2020MNRAS.496L..28L}. The number of bursts in each redshift range is plotted in Figure~\ref{fig:1}. This histogram could be well fitted by several distributions~\citep{2016PhRvL.117i1301M,2021ApJ...916...70Z,2022MNRAS.510L..18J,2021arXiv211107476Q}. Here, we fit the number density distribution by assuming that FRBs have a constant comoving number density,
\begin{equation}\label{Nconst}
N_{\text {const}}(z)=\mathcal{N}_{\text {const}} \frac{\chi^{2}(z)}{H(z)(1+z)} e^{-D_{\mathrm{L}}^2(z) /\left[2 D_{\mathrm{L}}^2\left(z_{\text {cut}}\right)\right]},
\end{equation}
where $D_{\mathrm{L}}$ is the luminosity distance, $\mathcal{N}_{\text {const}}$ is a normalization
factor to ensure that $N_{\text {const}}(z)$ integrates to unity, $z_{\text {cut}}$ is a Gaussian cutoff at some redshift to represent an instrumental signal-to-noise threshold. By confronting the histogram with Equation~\ref{Nconst}, we obtain that $z_{\mathrm{cut}}=0.8$ is consistent with the number density distribution. In the following analysis, we use this fitting results as the distribution for sampling the redshifts of upcoming FRBs.

\subsection{Lensing probability} \label{subsec:Lensing probability}
The lensing theory predicts that the probability of a source at redshift $z_s$ being lensed by a foreground dark matter halo is \citep{1992grle.book.....S}
\begin{equation}\label{lpro}
P=\int_{0}^{z_{s}} d z_{l} \frac{d D_{\mathrm{p}}}{d z_{l}} \int_{0}^{\infty} \sigma\left(M, z_{l}\right) n\left(M, z_{l}\right) d M,
\end{equation}
where $D_{\mathrm{p}}$ is the proper distance from the observer to the lens object,
$\sigma\left(M, z_{l}\right)$ is the lensing cross section at redshift $z_{l}$ for a dark halo of mass $M$,
$n\left(M, z_{l}\right)$ is the number density of dark matter halos with masses between $M$ and $M + dM$.
For a singular isothermal sphere(SIS) lens, the cross section producing two images with a flux ratio less than a given threshold $r$ is \citep{2002ApJ...566..652L}
\begin{equation}
\sigma(<r)=16 \pi^{3}\left(\frac{\sigma_{v}}{c}\right)^{4}\left(\frac{r-1}{r+1}\right)^{2}\left(\frac{D_{l} D_{l s}}{D_{s}}\right)^{2},
\end{equation}
where $\sigma_{v}$ is the velocity dispersion.
In addition, the number density of dark matter halos at redshift $z$ in the mass range from $M$ to $M + dM$ is
\begin{equation}
n(M, z) d M=\frac{\rho_{0}}{M} f(M, z) d M
\end{equation}
where $\rho_{0}=\Omega_{m} \rho_{\text {crit, } 0}$ is the mean mass density of the present universe, and $f(M, z)$ is the Press-Schechter function~\citep{1974ApJ...187..425P}. For a source at redshift $z_s$, we can calculate redshift $z_l$ which maximizes the lensing probability from Equation~\ref{lpro}. The relation between the redshift of a source and the one maximizing the lensing probability is shown in Figure 2 of \citet{2014SCPMA..57.1390L}. With sampled redshifts of FRB sources and the lens redshifts determined from the $z_s-z_l$ curve in Figure 2, we can mock the redshift information of near future lensed FRB systems for our following analysis. 

\subsection{Uncertainty estimation} \label{subsec:Uncertainty estimation}
For the time delay distance in Equation~\ref{Dt} and the kinetic distance in~\ref{Dd}, there are following five uncertainty sources: the measurement of the time delay, lens modeling for predicting $\Delta \phi$ and $J$, the velocity dispersion of the lensing galaxy, and contamination from masses along the line of sight (LOS).

For a typical galaxy-scale lensing system, the time delay between images is in the order of dozens of days, while the duration of an FRB is in the order of several milliseconds. Therefore, the relative uncertainty of time delay measurements between lensed FRB images roughly equals the ratio of pulse width ($\sim1$ ms) to time delay ($\sim10$ days), i.e. $10^{-3}/10^6\sim10^{-9}$. It suggests that time delay of this kind intriguing systems could be measured with extremely high precision and its relative uncertainty can be almost negligible.

The second ingredient of uncertainty is from the Fermat potential reconstruction and determined by lens modeling. For lensed transient systems, it is possible to reconstruct the mass distribution of the foreground lens galaxy more precisely due to the absence of dazzling AGN contamination in their source host galaxies. Simulations show that a cleaner image of the host galaxy without dazzling AGN in the center could improve the precision of Fermat potential reconstruction by a factor of 3~\citep{2017NatCo...8.1148L,2018NatCo...9.3833L}. In addition to this positive aspect, the mass-sheet degeneracy (MSD: a family  of mass density profiles could reproduce the same lensing observables, e.g. image positions and relative fluxes, but yield different measured values of cosmological parameters) is an intractable effect which might lead to reduce of precision and accuracy. Methods of breaking MSD have been intensively investigated and could be classified as two catagories, i.e. assuming theoretical priors~\citep{1997ApJ...490..493N,2020A&A...639A.101M,2020JCAP...11..045G,2021MNRAS.501..784D} and appealing non-lensing stellar kinematics data~\citep{2020A&A...643A.165B,2021A&A...649A..61B}.
More recently, \citet{2021MNRAS.504.5621D} carried out realistic simulations in lens modeling based on HST WFC3 observations from transient sources (e.g. supernovae, gamma-ray bursts, fast radio bursts, gravitational waves) to quantitatively compare the precision of $H_0$ inferred from lensed transient and the one from lensed AGN. It was found that the Shapiro delay and the geometric delay inferred from lensed transient systems are 3.8 and 4.7 times more precise than the ones from traditional lensed quasars, respectively. This result indicates that the precision of Fermat potential reconstruction could be improved by a factor of about 4. With all these investigations into account, it is reasonable to consider a 1.5\% relative uncertainty on Fermat potential reconstruction for upcoming lensed FRB systems~\citep{2021ApJ...916...70Z}. For the final simulation, as done in~\citet{2020MNRAS.497L..56Y}, we take the power-law lens model $\gamma^{\prime}$ to represent the error in $\Delta \phi$ and $J$ when the lens model is reconstructed.

For the uncertainty budget of stellar velocity dispersion, the Large Synoptic Survey Telescope (LSST) could provide
high-quality (sub-arcsecond) imaging data in general. Moreover, space-based facilities, such as HST and JWST, could also obtain additional images with high resolution for lensing systems. Meanwhile, the participation of other ground-based facilities makes it possible to derive additional spectroscopic information. Therefore, synergetic strategies will be helpful for coping with the fractional uncertainty of the stellar velocity dispersion. Here, consistent with \citet{2020MNRAS.497L..56Y}, we set 5\% as a reference for the average relative error of this budget. 

The last factor of uncertainty contribution for $D_{\Delta t}$ is LOS environment modeling. This budget is usually characterized by an external convergence ($\kappa_{\mathrm{ext}}$), which is resulted from the excess mass along the lines of sight to the lensing galaxies, and its effect on $D_{\Delta t}$ can be quantitatively expressed as $D_{\Delta t}=D_{\Delta t}^{\text {model}} /\left(1-\kappa_{\text {ext}}\right)$. Intensive and elaborate studies for some well-measured systems (i.e. HE 0435-1223), the fractional uncertainty caused by masses along the LOS could be controlled as less as 2.5\% through weighted galaxy counts~\citep{2002A&A...395...17W,2005Natur.435..629S,2011MNRAS.410.2167F,2017MNRAS.467.4220R}. Furthermore, sophisticated methods, such as the inpainting technique and multiscale entropy filtering algorithm, would yield a 1.6\% relative uncertainty of $\kappa_{\text {ext }}$ and thus the measurement of time delay distance \citep{2018MNRAS.477.5657T}. Therefore, we take a $2\%$ fractional uncertainty on average in LOS environment modeling for upcoming lensed FRB systems. 

The uncertainty levels of all budgets we adopted for the following simulation are summarized in Table \ref{tab:Parameters}. Then the total uncertainties of $D_{\Delta t}$ and $D_{l}$ can be propagated from corresponding budgets. It is obtained that typical uncertainties of $D_{\Delta t}$ and $D_{l}$ are at the level of 5\% and 11\%, respectively.

\begin{table}
	\centering
	\caption{Measurement uncertainty adopted for simulation}
	\label{tab:Parameters}
	\begin{tabular}{ccccc} 
		\hline
		\multicolumn{5}{c}{\text { Uncertainty }} \\
		\hline
		$\delta \Delta t / \Delta t$ & $\delta \gamma^{\prime}$ & astrometry (arcsec) & $\delta \sigma_{v} / \sigma_{v}$ & $\delta \kappa_{\text {ext}}$ \\
        0\% & 0.02 & 0.005 & 5\% & 2\% \\
		\hline
	\end{tabular}
\end{table}

\subsection{Statistical analysis and results}\label{subsec:Statistical analysis and results}
\begin{figure}
    \includegraphics[width=\columnwidth]{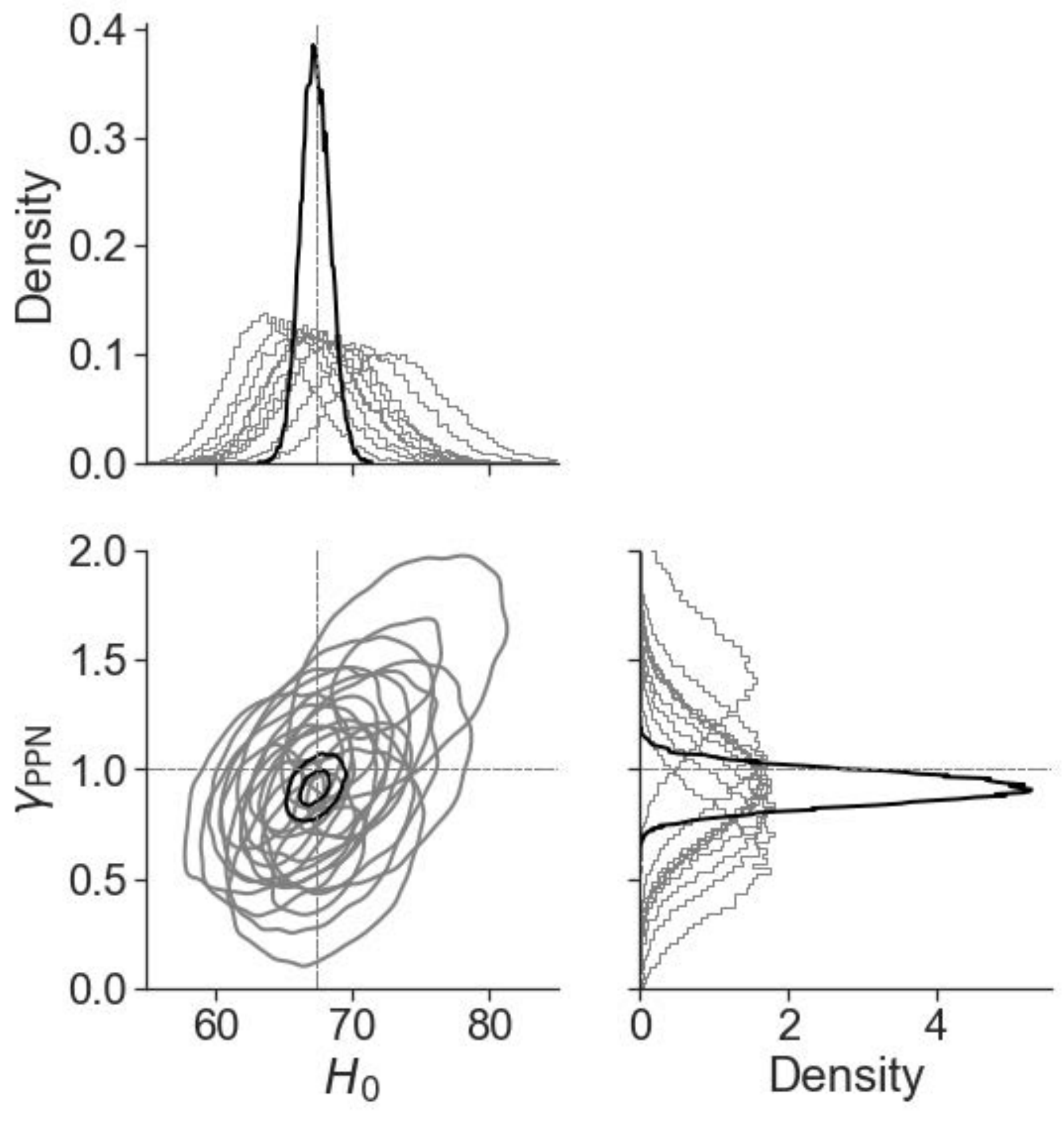}
    \caption{Simultaneous constraints on $H_0$ and $\gamma_{\mathrm{PPN}}$ from 10 strongly lensed FRB events. The black line is the result of 10 lenses, the gray line is the result of a single lens. Grey dashed lines denote the true values set in the fiducial model for simulation.
    \label{fig:2}}
\end{figure}

After setting up the above-mentioned concerning three aspects, we forecast the future constraining power on $H_0$ and $\gamma_{\text {PPN}}$ from 10 mocked strongly lened FRB systems. In the simulation, we use the fiducial model as GR and the latest $Planck$ $\Lambda$CDM cosmology with $H_{0}=67.4$ and $\Omega_{\mathrm m}=0.31$. We perform 5000 realizations to test whether simultaneous constraints on $H_0$ and $\gamma_{\text {PPN}}$ from lensed FRBs are biased. Results are presented in Figure \ref{fig:2}. It suggests that, in this context, $H_0$ could be bounded to (numerical result need to be included), corresponding to a 1.5\% precision. This is $>3$ times better than the one constrained from four well-measured lensed quasars and almost comparable with the one from simulated future 40 lensed quasar systems~\citep{2020MNRAS.497L..56Y}. Meanwhile, the precision of this simultaneous constraint is even about 2 times better than one obtained in~\citet{2019MNRAS.484.4726B} where only one parameter $H_0$ was estimated using posteriors of time delay distances of four well-measured lensed quasars. 
For $\gamma_{\mathrm{PPN}}$, it is constrained to (numerical result need to be included), corresponding to a 8.7\% precision. This is $>2$ times better than the one constrained from four currently available lensed quasars and almost comparable with the one from simulated future 40 lensed quasar systems~\citep{2020MNRAS.497L..56Y}. Moreover, simultaneous constraints on $H_0$ and $\gamma_{\mathrm{PPN}}$ are in excellent agreement with the input values of these two parameters in the fiducial model used for simulation. It implies that the proposed method is valid to achieve unbiased joint estimation for $H_0$ and $\gamma_{\mathrm{PPN}}$ from strongly lensed FRB systems. It can also be foreseen that strongly lensed FRB systems have a promising prospect to simultaneously constrain these two fundamental parameters.

\section{CONCLUSIONS}\label{sec:conclusions}
In this work, we first investigate the possibility of simultaneously estimating Hubble constant $H_0$ and post-Newtonian parameter $\gamma_{\mathrm {PPN}}$ with strongly lensed FRB systems. Combining measurements of time delay, stellar velocity dispersion, high resolution images, and LOS environment modeling, we simultaneously obtain a precision of $\sim 1.5\%$ on $H_0$ and constrain $\gamma_{\mathrm {PPN}}$ to a precision of  $\sim 8.7\%$ from 10 lensed FRBs systems which are likely to achieve in several years. In view of the Hubble constant tension and foundation of GR as gravity theory, these simultaneous determinations, will be useful for properly revealing possible degeneracy between these two fundamental issues.

In the following several years, wide-field sensitive radio telescopes, such as the CHIME/FRB instruments, would probably detect a considerable population of repeating FRBs~\citep{2019Natur.566..235C}. For these repeaters, it is likely to localize them to their host galaxies with very long baseline array and then obtain their redshift information. More promisingly, even though for apparent one-off FRBs, powerful facilities including radio interferometers, world-leading radio survey telescopes, and multi-messenger discovery engines for the next decade, could detect and localize $\sim 10^4$ this kind of events each year~\citep{2019BAAS...51g.255H}. It is predictable that, with the rapid progress in FRB community, strongly lensed FRBs will be competitive to existing strongly lensed quasars in simultaneously determining the cosmic expansion rate and testing the validity of general relativity. 

\vspace{2.5\parskip}

\section{Acknowledgements}
We would like to thank Tao Yang for helpful discussions. This work was supported by the National Key Research and Development Program of China Grant No. 2021YFC2203001, the National Natural Science Foundation of China under Grants Nos. 11920101003, 11722324, 11603003, 11633001, and U1831122, the science research grants from the China Manned Space Project with No. CMS-CSST-2021-B11, the Strategic Priority Research Program of the Chinese Academy of Sciences, Grant No. XDB23040100, and the Interdiscipline Research Funds of Beijing Normal University.

\section*{Data Availability}
The data underlying this article will be shared on reasonable request to the corresponding author.

\bibliography{refs}{}
\bibliographystyle{mnras}

\bsp	
\label{lastpage}
\end{document}